\documentclass{appolb}
\usepackage{graphicx}
\usepackage[caption=false]{subfig}
\usepackage{amsmath}
\usepackage{float}
\usepackage[hidelinks]{hyperref}

\begin{document}
\title{nuSTORM: a unique GeV electron-(anti)neutrino machine\thanks{Presented at the Ultra-Cold Muons for the LHC Workshop, September 2026.}}

\author{Xianguo Lu\address{Department of Physics, University of Warwick, Coventry CV4 7AL, UK}}
\maketitle

\begin{abstract}

In this conference contribution, I briefly review  current experimental programmes measuring the  Dirac CP phase $\delta_\textrm{CP}$ using accelerator neutrinos. I argue that the nuSTORM concept---neutrino beams from a muon storage ring---could provide high-quality electron-(anti)neutrino beams needed for the cross-section inputs to $\delta_\textrm{CP}$ measurements. 
   
\end{abstract}

\section{Introduction}

In the Standard Model, 
neutrinos are identified 
by the flavours of their charged-lepton partners in charged-current interactions. 
Neutrinos are observed to have non-zero masses,  requiring extension beyond the Standard Model. 
Their mass eigenstates are stationary states in vacuum. The flavour and mass eigenstates can be transformed into each other via the Pontecorvo--Maki--Nakagawa--Sakata (PMNS) matrix, 
\begin{align}
|\nu_\alpha\rangle&=\sum_{i=1}^{3}U_{\alpha i}^{*}|\nu_i\rangle,
\end{align}
where the indices $\alpha=e,\mu,\tau$  and $i=1,2,3$ are for the flavour and mass eigenstates, respectively, and $U_{\alpha i}$ are the PMNS matrix elements. The non-zero mass-squared splittings
\begin{align}
\Delta m^2_{ij}=m_i^2-m_j^2 
\end{align}
cause the non-stationary flavour eigenstates to oscillate.

The three different masses introduce the problem of neutrino mass ordering (NMO): the ordering is normal if the intermediate mass lies closer to the lightest one, and inverted if closer to the heaviest---only a problem when there are more than two neutrino masses. By definition, the mass-squared splitting between mass eigenstates 2 and 1 is the  solar splitting~\cite{ParticleDataGroup:2026mpi} 
\begin{align}
    \Delta m^2_{21}\simeq7.5\times10^{-5}\,\mathrm{eV}^{2}.\label{eq:dm12}
\end{align}
The splitting between mass eigenstates 3 and 2 is the atmospheric splitting, which is positive for normal ordering ($m_1<m_2<m_3$) and negative for inverted ordering ($m_3<m_1<m_2$)~\cite{ParticleDataGroup:2026mpi}:
\begin{align}
    \Delta m^2_{32}\simeq\pm2.5\times10^{-3}\,\mathrm{eV}^{2}, \label{eq:dm23}
\end{align}
 approximately 30 times that of the solar splitting. These two independent mass-squared  splittings result in two characteristic oscillation scales in  $L/E$, 
where $L$ is the propagation distance (baseline) and $E$ is the neutrino energy.

In the standard parameterisation, the PMNS matrix, omitting the Majorana phases that do not affect oscillations, is given by
\begin{equation}
\label{eq:pmns_matrix}
\begin{gathered}
\resizebox{0.88\columnwidth}{!}{$\displaystyle
U=
\begin{pmatrix}
1&0&0\\
0&c_{23}&s_{23}\\
0&-s_{23}&c_{23}
\end{pmatrix}
\begin{pmatrix}
c_{13}&0&s_{13}e^{-i\delta_{\mathrm{CP}}}\\
0&1&0\\
-s_{13}e^{i\delta_{\mathrm{CP}}}&0&c_{13}
\end{pmatrix}
\begin{pmatrix}
c_{12}&s_{12}&0\\
-s_{12}&c_{12}&0\\
0&0&1
\end{pmatrix}
$}
\end{gathered}
\end{equation}
where $s_{ij}=\sin\theta_{ij}$ and $c_{ij}=\cos\theta_{ij}$, with the mixing angles $\theta_{12}$, $\theta_{13}$ and $\theta_{23}$ and the Dirac CP phase $\delta_{\mathrm{CP}}$. The angle $\theta_{12}$ rotates the 1--2 sector, while $\theta_{23}$ mixes the $\mu$- and $\tau$-flavour eigenstates: a maximal mixing ($\theta_{23}=\pi/4$) could suggest an underlying symmetry involving the $\mu$ and $\tau$ flavours. If $\theta_{13}$ were zero,  $\delta_{\mathrm{CP}}$ would be inaccessible. 

Current neutrino-oscillation programmes   aim to determine both the mass and flavour structures of  neutrinos: the former through precision measurements of the mass-squared splittings and a determination of NMO, and the latter through measurements of the mixing angles and  $\delta_{\mathrm{CP}}$. 
   Other open questions remain beyond the reach of these programmes, including the absolute neutrino masses and whether neutrinos are  Dirac or Majorana particles. This contribution briefly reviews  $\delta_\textrm{CP}$ measurements with accelerator neutrinos and (re)introduces the nuSTORM concept as a unique way to improve measurements of electron-(anti)neutrino\footnote{The physical case extends to both neutrinos and antineutrinos. In the following, ``anti'' is omitted wherever no confusion is caused.} interactions.   

\section{Neutrino CP violation}

The three neutrino masses give two independent mass-squared splittings, which, together with the factor $L/E$, determine the  oscillation phases in vacuum:  
\begin{equation}
\Delta_{ij}=\frac{\Delta m^2_{ij}L}{4E}
\simeq1.267\,\frac{\Delta m^2_{ij}[\mathrm{eV}^2]\,L[\mathrm{km}]}{E[\mathrm{GeV}]}.
\end{equation}
The characteristic oscillation scale---in terms of $L/E$---driven by the atmospheric mass-squared splitting [Eq.~(\ref{eq:dm23})] is therefore in the $\mathrm{km/MeV}$ range. 
Oscillations at this scale have been observed with reactor neutrinos, accelerator neutrinos, and atmospheric neutrinos, as summarised in Table~\ref{tab:atmospheric_experiments}.

\begin{table}[!hptb]
\centering
\caption{Representative neutrino energies and baselines in   experiments measuring oscillations predominantly driven by the atmospheric mass-squared splitting. The flux column shows the flavours at source for the signal samples. }
\label{tab:atmospheric_experiments}
\begin{tabular}{@{}llcc@{}}
\hline
Experiment & Flux & $E$ & $L$ \\
\hline
Daya Bay~\cite{DayaBay:2012fng} &  $\bar\nu_e$ & $2$--$8$ MeV & $1.65$ km \\
RENO~\cite{RENO:2012mkc} &  $\bar\nu_e$ & $2$--$8$ MeV & $1.44$ km \\
Double Chooz~\cite{DoubleChooz:2011ymz} &  $\bar\nu_e$ & $2$--$8$ MeV & $1.05$ km \\
T2K~\cite{T2K:2011ypd,T2K:2012qoq} &  $\nu_\mu,\bar\nu_\mu$ & $\sim0.6$ GeV & $295$ km \\
NOvA~\cite{NOvA:2016kwd,NOvA:2016vij} &  $\nu_\mu,\bar\nu_\mu$ & $\sim2$ GeV & $810$ km \\
MINOS~\cite{MINOS:2006foh} &  $\nu_\mu,\bar\nu_\mu$ & $\sim3$ GeV & $735$ km \\
Super-Kamiokande~\cite{Super-Kamiokande:1998kpq} &  $\nu_\mu,\bar\nu_\mu,\nu_e,\bar\nu_e$ & $\sim0.1$--$100$ GeV & $\sim10$--$13\,000$ km \\
IceCube/DeepCore~\cite{IceCube:2017lak} &  $\nu_\mu,\bar\nu_\mu$ & $\sim5$--$50$ GeV & $\sim10$--$13\,000$ km \\
\hline
\end{tabular}
\end{table}

The oscillations primarily driven by the solar mass-squared splitting [Eq.~(\ref{eq:dm12})] have an $L/E$-scale approximately 30 times larger, and therefore are mainly observed in reactor experiments\footnote{The solar mass-squared splitting has also been probed with solar neutrinos through matter effects in the sun~\cite{SNO:2011hxd}.}~\cite{KamLAND:2002uet, SNO:2025chx, JUNO:2025gmd}. With a terrestrial baseline, neutrino energies at oscillation experiments are limited up to a few tens of GeV. 

With an accelerator muon-neutrino beam, both disappearance and electron-neutrino appearance can be observed, but  the latter gives the dominant sensitivity to $\delta_\mathrm{CP}$. 
  The difference between the neutrino and antineutrino appearance probabilities in vacuum is 
\begin{equation}
\begin{aligned}
P_{\nu_\mu\to \nu_e}-P_{\bar\nu_\mu\to\bar\nu_e}
&=-16J_{\mathrm{CP}}\sin\Delta_{21}\sin\Delta_{31}\sin\Delta_{32},\\
J_{\mathrm{CP}}&=c_{12}s_{12}c_{23}s_{23}c_{13}^{2}s_{13}
\sin\delta_{\mathrm{CP}},
\end{aligned}
\label{eq:cpasym}
\end{equation}
where $J_{\mathrm{CP}}$ is the Jarlskog invariant measuring the strength of the intrinsic CP violation. The measured mixing angles give $J_{\mathrm{CP}}\simeq 0.03\sin\delta_{\mathrm{CP}}$~\cite{ParticleDataGroup:2026mpi}. Currently, accelerator neutrino experiments~\cite{T2K:2025wet} provide the dominant constraints on $\delta_{\mathrm{CP}}$, while atmospheric neutrino measurements~\cite{Super-Kamiokande:2023ahc} also provide sensitivity. Hyper-Kamiokande~\cite{Hyper-Kamiokande:2025fci} and DUNE~\cite{DUNE:2020jqi} aim to measure $\delta_{\mathrm{CP}}$ with greater precision. 

 \section{Accelerator neutrinos}
 
\begin{figure}[!hptb]
\centering
\includegraphics[width=0.70\columnwidth]{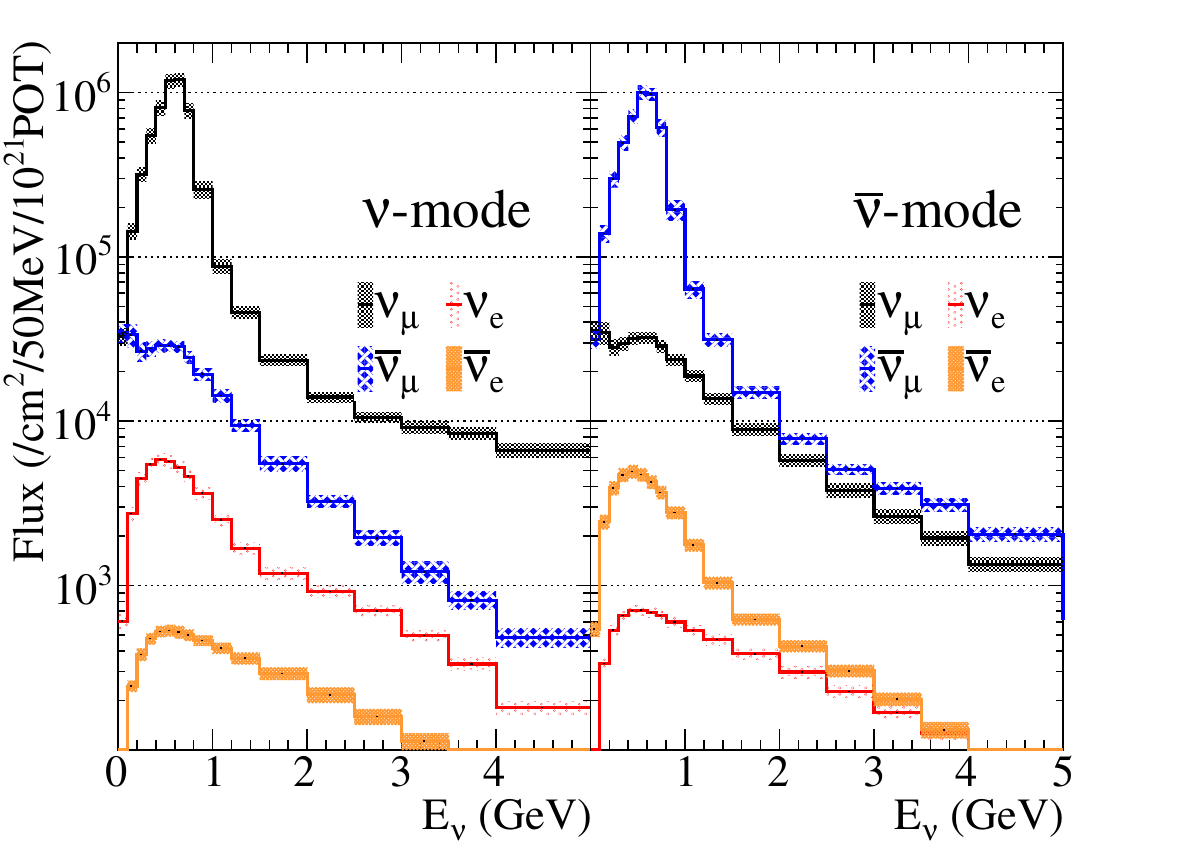}
\caption{Predicted unoscillated neutrino fluxes at
Super-Kamiokande---the far detector of T2K. Taken from~\cite{T2K:2015gcs}.}
\label{fig:t2k_beam}
\end{figure}

In all the accelerator neutrino experiments mentioned above, the neutrino beams are generated via secondary hadron decays in flight following a proton beam bombarding a target.  The dominant contributions to the $\nu_\mu$ and $\bar\nu_\mu$ beams are
\begin{equation}
\pi^+\rightarrow\mu^++\nu_\mu,\qquad 
\pi^-\rightarrow\mu^-+\bar\nu_\mu,
\end{equation}
respectively. Kaon and muon decays also contribute to the electron-neutrino flux  components.  
Downstream magnetic horns focus the secondary hadrons of one charge and defocus the other, producing a neutrino-enhanced or antineutrino-enhanced beam (Fig.~\ref{fig:t2k_beam}). Following the horn is an uninstrumented volume, the decay pipe, where most of the decays happen.

Near detectors are used to constrain the neutrino flux- and interaction-model uncertainties, while far detectors measure the oscillated neutrino spectra to determine the oscillation parameters. In conventional hadron-decay neutrino beams, flux uncertainty arises from hadron production and reinteractions, as well as uncertainties in the proton beam and the horn system~\cite{T2K:2012bge}. The intrinsic electron-neutrino components remain irreducible backgrounds to the appearance measurements.

\subsection{Electron-neutrino interactions for $\delta_\textrm{CP}$ measurements}

\begin{figure}[!hptb]
\centering
\includegraphics[width=0.7\columnwidth]{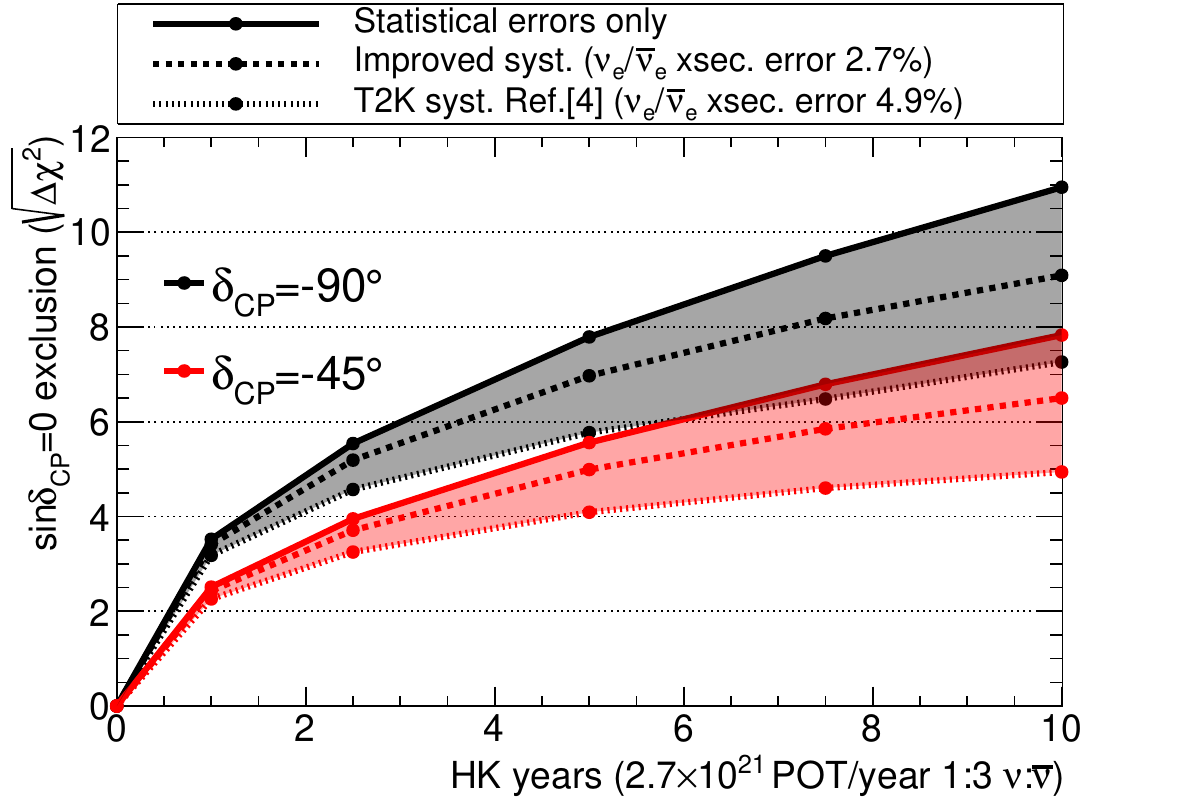}
\caption{Projected CP-violation discovery sensitivity at Hyper-Kamiokande with  different assumptions of systematic uncertainties. See text. Taken from~\cite{Hyper-Kamiokande:2025fci}.}
\label{fig:hyperk_cp}
\end{figure}

In  $\delta_\textrm{CP}$ measurements, electron-neutrino cross sections are identified as a major source of systematic uncertainties~\cite{Hyper-Kamiokande:2025fci,DUNE:2020jqi}. For example,  for a non-maximal CP-violating phase of  $\delta_\textrm{CP}=-45^\circ$ (assuming NMO is known), a 5$\sigma$ discovery sensitivity of CP violation at Hyper-Kamiokande could be achieved in approximately five years under the  scenario with improved systematic uncertainties, rather than approximately  ten years with T2K-like systematic uncertainties (Fig.~\ref{fig:hyperk_cp}). In particular, the assumed uncertainty on
the $\nu_e/\bar\nu_e$ cross-section ratio is reduced from 4.9\% to 2.7\% in the improved scenario~\cite{Hyper-Kamiokande:2025fci}.

A major challenge is the limited precision of existing cross-section data for electron neutrinos at the relevant  energies and on the relevant targets. 
In a $\delta_\textrm{CP}$ measurement using conventional hadron-decay beams, the electron-neutrino beam components are suppressed at relevant energies, as they are intrinsic  backgrounds to the appearance measurements at the far detector. As a result, near-detector samples are overwhelmed by  muon-neutrino interactions, which can contaminate the electron-neutrino  signals through particle misidentification. This imposes stringent particle-identification and background-rejection requirements on \textit{in situ} electron-neutrino cross-section measurements.  
Lepton-flavour universality relates electron- and muon-flavour charged-current interactions at the vertex level, but extrapolating muon-neutrino cross sections to electron neutrinos requires accounting for charged-lepton mass effects, radiative corrections~\cite{Tomalak:2021hec} and their interplay with nuclear effects. 

Further improving electron-neutrino cross sections,  therefore, requires dedicated  electron-neutrino  beams, which should provide
\begin{enumerate}
    \item neutrino energies relevant to electron-neutrino appearance;
    \item  well-controlled fluxes;
    \item  high statistics; and 
    \item low backgrounds from muon-neutrino interactions.
\end{enumerate}
     
\section{Neutrinos from STORed Muons}

\begin{figure}[!hptb]
\centering
\includegraphics[width=0.99\columnwidth]{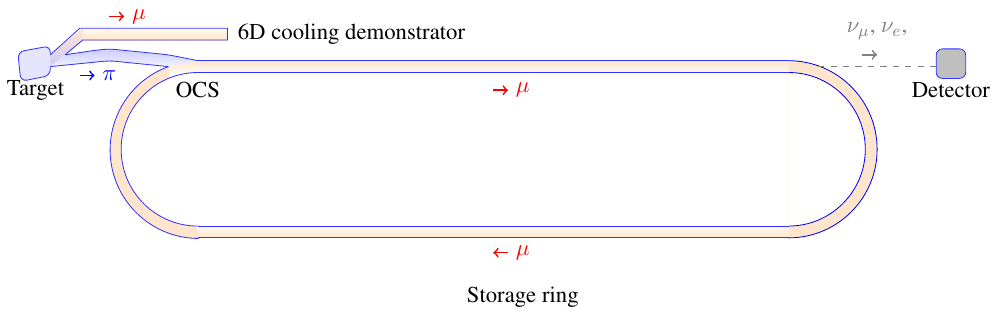}
\caption{Schematic of the nuSTORM facility. See text. Taken from~\cite{nuSTORM:2022div}.}
\label{fig:nustorm_ring}
\end{figure}

The nuSTORM (Neutrinos from STORed Muons) facility~\cite{nuSTORM:2013cqr, Adey:2013afh, nuSTORM:2012jbd, Ahdida:2020whw, nuSTORM:2022div,nuSTORM:2025tph} is designed to produce neutrino beams from muon decays in flight,
\begin{equation}
\begin{aligned}
\mu^+&\rightarrow e^++\nu_e+\bar\nu_\mu,\\
\mu^-&\rightarrow e^-+\bar\nu_e+\nu_\mu,
\end{aligned}
\label{eq:muon_decay}
\end{equation}
in a storage ring, where the neutrino parents can be precisely monitored, as opposed to those in the decay pipe in a conventional hadron-decay beam. With circulating $\mu^{+}$ ($\mu^{-}$), the beam is a  $1:1$  mixture of $\nu_e$ and $\bar\nu_\mu$ ($\bar\nu_e$ and $\nu_\mu$) at production. The facility consists of a target, a pion transport line, a racetrack-shaped muon storage ring, and a detector, as shown in Fig.~\ref{fig:nustorm_ring}. 

Secondary pions produced by the proton beam striking the target are focused by a magnetic horn and transported through a charge-and-momentum-selective transfer line into the storage ring. Pions decaying in the production straight produce muons and a prompt neutrino pulse, known as the pion flash, which is a separable beam component in time. The production straight uses a focusing–defocusing lattice to transport the injected pions and is optimised for the capture efficiency of the daughter muons. 

The captured muons circulate for a few tens of turns, through the production straight, the return arcs and the connecting return straight. Muons decaying in the production straight produce a neutrino beam towards the detector downstream. The return arcs and return straight use a fixed-field alternating-gradient lattice that provides a large momentum acceptance. The ring is designed to operate at a central momentum between 1 and 6~GeV/$c$ with a momentum acceptance of up to $\pm16\%$.

Dedicated ring instrumentation would measure the circulating muon intensity, momentum, and beam properties, including the orbit and profile, 
providing precise monitoring of the stored muons, and hence the shape (thanks to the lepton-decay kinematics) and normalisation of the neutrino flux. Because the circulating muons---the neutrino parents---are controlled and monitored, the upstream uncertainties from  the proton beam, the hadron production and reinteractions, and the pion transport line (including the horn system) do not propagate. The design aims to achieve a normalisation uncertainty of the neutrino  flux  much below 1\%.

With suitable detectors and relevant targets, nuSTORM could therefore deliver a neutrino cross-section programme with negligible systematic uncertainties from flux normalisation and shape for both muon  and electron neutrinos.  While  $\nu_e$ and $\bar\nu_\mu$ ($\bar\nu_e$ and $\nu_\mu$) are produced in equal amounts at nuSTORM, the electron-neutrino samples are no longer dominated by the muon-neutrino background. The neutrino energy and flux can be optimised via the pion momentum selection and the stored-muon momentum and ring acceptance, in principle providing high-statistics samples at the desired  energies relevant for $\delta_\mathrm{CP}$ measurements.

\section{Summary}

One of the major open questions in particle physics is neutrino CP violation. The Dirac CP phase $\delta_\textrm{CP}$ is one of the fundamental parameters beyond the Standard Model. Current  experimental programmes measuring $\delta_\textrm{CP}$ rely on the electron-neutrino appearance in muon-neutrino beams at energies in the sub-GeV to few-GeV regime, which in turn requires precise modelling of electron-neutrino cross sections as an input. The nuSTORM concept---neutrino beams from a muon storage ring---could potentially provide  dedicated measurements of neutrino interactions relevant for neutrino oscillations, in particular with its uniquely high-quality electron-neutrino beams. Because of the unique control of the neutrino parents via monitoring the stored muons, the nuSTORM flux is designed to achieve a normalisation uncertainty much better than 1\% with a precisely determined shape. 
nuSTORM  could therefore contribute uniquely to the $\delta_\textrm{CP}$ programme, complementing conventional hadron-decay neutrino beams.

\bibliographystyle{unsrt}
\bibliography{references}

\end{document}